# Two-dimensional gridless super-resolution method for ISAR imaging

Mohammad Roueinfar [a], Mohammad Hossein Kahaei [a*]

[a] School of Electrical Engineering, Iran University of Science and Technology, Tehran, Iran

**Abstract:** We are focused on improving the resolution of images of moving targets in Inverse Synthetic Aperture Radar (ISAR) imaging. This could be achieved by recovering the scattering points of a target that have stronger reflections than other target points, leading to increasing the higher Radar Cross Section (RCS) of a target. These points, however, are sparse and when the received data is incomplete, moving targets would not be properly recognizable in ISAR images. To increase the resolution in ISAR imaging, we propose the 2-Dimensional-Reweighted Trace Minimization (2D-RWTM) method to retrieve frequencies of sparse scattering points in both range and cross-range directions. This method is a gridless super-resolution method, which does not depend on fitting the scattering point on the grids, leading to less complexity compared to the other methods. Using computer simulations, the proposed 2D-RWTM is compared to the Atomic Norm Minimization (ANM) in terms of the Mean Squared Errors (MSE). The results show that using the proposed method, the scattering points of a target are successfully recovered. It is shown that by selecting different weighting matrices and scattering points adjacent to each other, the recovery in ISAR imaging is still successful.

## 1 Introduction

ISAR is an imaging radar for moving targets, which can form a two-dimensional image of a target by adding a second dimension as a cross-section and benefiting from dual processing gain for range and cross-range direction[1]. This structure, wherein the radar is fixed and the targets are moving, can identify and distinguish moving targets from each other[2].

In 2D ISAR imaging cross-range direction, *i.e.*, the second dimension, is obtained based on the relative motion between the radar antenna and the target. Then, a 2D image is obtained by collecting the scattering points of a target from different viewing angles that are generated by the relative motion between the radar and the target[3]. It is well known that the resolution in the direction of the range depends on the bandwidth of the transmitted signal and the cross-range direction is determined by the Coherent Processing Interval (CPI). However, due to the limitations in the transmitted signal bandwidth and the CPI, it is not always possible to increase the resolution in each direction. In order to overcome these limitations, some super-resolution methods have been introduced to increase the resolution with no need for increasing the signal bandwidth and the CPI[3]. These methods can be categorized as

- Higher-resolution spectrum estimation[6-13],
- Sparse Bayesian Learning (SBL)[14-16],
- Compressive sensing (CS)[3-5], and
- Gridless sparse methods[17].

Higher-resolution spectrum estimation methods often require some initial information such as the number of sources and the Nyquist rate sampling[5]. The SBL methods requires prior information such as probability distribution of previously observed data[21].

Due to the sparse nature of the target scattering points, CS methods have been developed[3-5]. The goal of these methods is to recover unknown scattering points of a target from a limited number of samples. These points generate strong RCSs at certain frequencies and the more are retrieved, the higher resolution in ISAR images is achieved. However, a major difficulty with the CS approach is that these methods are designed based on a predefined grid, which incorporates a discrete sparse dictionary into the computations that would be in practice incompatible with the scattering points. In fact, as the target location is not predictable, the scattering points with high RCSs would not necessarily place on these grids. In this case, to get higher resolution, it is necessary to increase the number of grids, which leads to more complexity of the algorithm.

To cope with this problem, as an alternative, gridless-based methods have been introduced. These methods may be interpreted as a continuous version of CS approaches, whose solution involve convex optimization problems by using Atomic-Norm Minimization (ANM). Gridless-based methods are faster and less complex than the other methods as they do not require signal discretization[18-19]. A gridless based method referred to as the Reweighted Atomic-norm Minimization (RAM) has been introduced in Ref. 20, which has accordingly been developed for ISAR imaging in Ref. 17. This algorithm executes a convex optimization problem based on ANM using an appropriate weighting matrix. It has been shown that RAM can enhance the sparsity and thus the resolution. In Ref. 22, ANM has been expanded for transmitting Frequency-Stepped Chirp Signals (FSCS), which is solved by an SDP. Then, the scattering points are recovered using off-the-shelf SDP solvers. In this work, we propose the 2D-Reweighted Trace Minimization (2D-RWTM) method for ISAR imaging by retrieving the 2D-frequencies of sparse scattering points of a target and show its effectiveness in enhancing the resolution of ISAR images.

The paper is organized as follows. In Sec. 2, the signal model is given. Sec. 3 is devoted to the proposed 2D-RWTM method and Sec. 4 presents the experimental results. Finally, Sec. 5 concludes the paper.

## 2 Signal Model

According to the principles of ISAR imaging, the relative movement between radar and a target can lead to two types of motion: *a) translational* and b) *rotational.* The transitional translational motion occurs when the entire target body is displaced, which generally leads to darkening. But in the other type of motion, only rotational and manoeuvring motions of a target in its place are considered. Here, we assume that the translational and the rotational motions are compensated. There are different methods for motion compensation in ISAR imaging, but in practice the latter assumption is achievable, which facilitates reducing the computational complexity. However, a major problem with ISAR images is that they are basically reconstructed from the pulses received from different angles of a target, which for any possible reasons may be missed during transmission, leading to blurred images. In Fig. 1, an ISAR radar observes only the signals from the angles $\theta_0, \theta_2,$ and $\theta_4$, out of 6 viewing angles and misses the others.

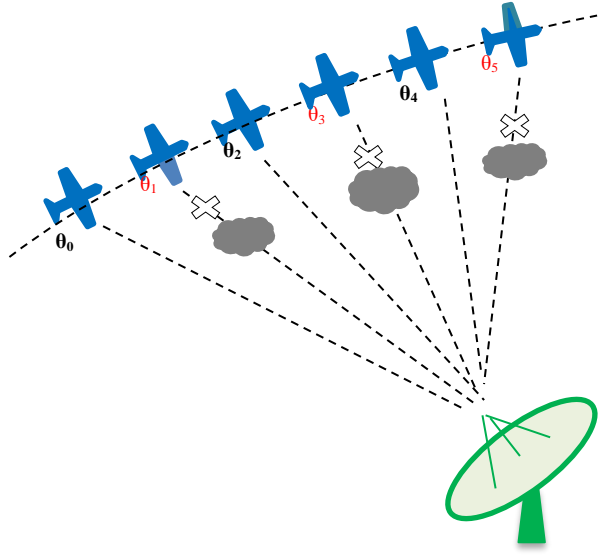

**Fig. 1**. ISAR radar receiving the signals incompletely.

Also, we consider the FSCS, which has a relatively low instantaneous bandwidth compared to the total processing bandwidth. This signal is made up of smaller and discrete components called bursts, each one consisting of a number of pulses modulated by Linear Frequency Modulation (LFM). It is assumed that the target is composed of $K$ scattering points with coefficients $\sigma_k$, which have high RCS. Then, the received signal after dechirping is obtained as

$$S_R(\theta_n, f_m) = \sum_{k=0}^{K-1} \sigma_k \exp\left[-j2\pi f_m \frac{2(R_0 - x_k \sin(\theta_n) + y_k \cos(\theta_n))}{c}\right], \quad (1)$$

where $f_m$ is the central frequency of the $m$-th pulse, $\theta_n$ presents the $n$-th viewing angle, $(x_k, y_k)$ show the coordinates of the $k$-th scattering point, $R_0$ is the instantaneous distance between the radar and target, and $c$ is the light speed. Assuming a small viewing angle, which leads to the approximations $(\theta_n) \approx \theta_n$, $\cos(\theta_n) \approx 1$, and compensating for the translational motion, the received signal is simplified as

$$S_R(\theta_n, f_m) = \sum_{k=0}^{K-1} \sigma_k \exp\left[-j4\pi f_m \frac{x_k \theta_n}{c}\right] \exp\left[-j4\pi f_m \frac{y_k}{c}\right]. \quad (2)$$

By defining $f_{x,n} = 4\pi f_m \theta_n/c$ and $f_{y,m} = 4\pi f_m/c$ in Eq. (2), and according to $f_m = f_0 + m\Delta f$ in a FSCS signal, where $f_0$ is the fundamental carrier frequency in a burst and $\Delta f$ is the frequency step for each pulse relative to the next pulse and assuming $f_0$ is larger than $\Delta f$ and $\theta_n$, we have $f_{x,n} \approx 4\pi f_0 \theta_n/c$. Thus, Eq. (2) can be written as

$$S_R(\theta_n, f_m) = \sum_{k=0}^{K} \sigma_k \exp[j(f_{x,n} x_k + f_{y,m} y_k)]. \tag{3}$$

Also, considering $N$ viewing angles and $M$ pulses in a burst of the transmitted signal, we have different values of $m$ and $n$ for $f_{x,n}$ and $f_{y,m}$, and we can present the received signal in matrix form as

$$\boldsymbol{S} = \begin{bmatrix} S_R(\theta_0, f_0) & \cdots & S_R(\theta_0, f_{M-1}) \\ \vdots & \ddots & \vdots \\ S_R(\theta_{N-1}, f_0) & \cdots & S_R(\theta_{N-1}, f_{M-1}) \end{bmatrix} \in \mathbb{C}^{N \times M} \tag{4}$$

or in vector form as

$$\boldsymbol{s}_{nm} = \text{vec}(\boldsymbol{S}^T) \in \mathbb{C}^{NM \times 1}. \tag{5}$$

## 3 2D-RWTM Algorithm

We propose an ISAR imaging algorithm based on the 2D-RWTM algorithm[18] in order to increase the resolution of ISAR images by retrieving the scattering points of a target in two dimensions. To do so, Eq. (3) can be presented using the Kronecker product of two-dimensional atoms in the direction of range and cross-range as

$$\boldsymbol{s}_{n,m} = \sum_{k=1}^{K} \sigma_k \, \boldsymbol{a}(f_{x,i}) \otimes \boldsymbol{a}(f_{y,i}). \tag{6}$$

where $\otimes$ shows the Kronecker product, $\sigma_k$ is the RCS of scattering points, and $K$ is the number of scattering points. Also, $\boldsymbol{a}(f_{x,i})$ and $\boldsymbol{a}(f_{y,i})$ are the atoms in the direction of range and cross-range, respectively defined as

$$\boldsymbol{a}(f_{x,i}) = \frac{1}{N}[1, \exp(-jf_{x,i}), \ldots, \exp(-j(N-1)f_{x,i})]^T, \tag{7}$$

$$\boldsymbol{a}(f_{y,i}) = \frac{1}{M}[1, \exp(-jf_{y,i}), \ldots, \exp(-j(M-1)f_{y,i})]^T. \tag{8}$$

Also, the received signal vector can be shown as

$$\boldsymbol{s}_{n,m} = \sum_{i=1}^{K} \sigma_k \boldsymbol{c}(\boldsymbol{f}_i), \tag{9}$$

where $\boldsymbol{f}_i = (f_{x,i}, f_{y,i})$ showing a 2D frequency for the i-th scattering point and $\boldsymbol{c}(\boldsymbol{f}_i) = \boldsymbol{a}(f_{x,i}) \otimes \boldsymbol{a}(f_{y,i}) \in \mathbb{C}^{NM \times 1}$ is a 2D normalized complex atom.

$$\mathcal{A} = \{\boldsymbol{c}(\boldsymbol{f}_i) | \boldsymbol{f}_i \in [0,1] \times [0,1]\}, \tag{10}$$

In other words, the received signal is a linear combination of the atoms of $\mathcal{A}$. Then, based on $\mathcal{A}$, the sparse scattering points of a target received by ISAR can be recovered by solving the following ANM optimization problem,

$$\|\boldsymbol{s}_{n,m}\|_{\mathcal{A}} = \inf_{\boldsymbol{f}_i \in [0,1] \times [0,1]} \left\{ \sum_i |\sigma_k| \,\middle|\, \boldsymbol{s}_{n,m} = \sum_{i=1}^{K} \sigma_k \boldsymbol{c}(\boldsymbol{f}_i) \right\}. \tag{11}$$

Then, the sparse scattering points of a target is recovered as

$$\hat{\boldsymbol{s}}_{n,m} = \arg\min_{\boldsymbol{s}_{n,m}^*} \|\boldsymbol{s}_{n,m}\|_{\mathcal{A}} \qquad \text{subject to} \quad \boldsymbol{S} = \boldsymbol{S}_{\Omega}, \tag{12}$$

where $\boldsymbol{S}$ is the complete data matrix (or full observations) as given by Eq. (4), $\boldsymbol{S}_{\Omega}$ shows the measured (or incomplete) data matrix, and $\Omega \in \{0, \ldots, N-1\} \times \{0, \ldots, M-1\}$ is a subset of the pulse and viewing angle indices. For example, in Fig.1, with 6 viewing angles and 8 pulses for each viewing angle (N=6, M=8) and 3 missing signals at $\theta_1, \theta_3,$ and $\theta_5$, and the missing frequencies $f_1, f_3, f_4, f_5, f_7$ of the 1th, 3th, 4th, 5th and 7th pulse at each viewing angle, $\boldsymbol{S}_{\Omega}$ is represented as

$$S_\Omega = \begin{bmatrix} S_{00} & 0 & S_{02} & 0 & 0 & 0 & S_{06} & 0 \\ 0 & 0 & 0 & 0 & 0 & 0 & 0 & 0 \\ S_{20} & 0 & S_{22} & 0 & 0 & 0 & S_{26} & 0 \\ 0 & 0 & 0 & 0 & 0 & 0 & 0 & 0 \\ S_{40} & 0 & S_{42} & 0 & 0 & 0 & S_{46} & 0 \\ 0 & 0 & 0 & 0 & 0 & 0 & 0 & 0 \end{bmatrix}$$

where $S_{nm} = S_R(\theta_n, f_m)$.

Due to the complexity of solving Eq. (12), an approximated solution based on Semidefinite Programing (SDP) is used. This can be done using the data matrix and defining the 2-Level Toeplitz (2LT) matrices[18-19]. In this work, we apply the Vandermonde decomposition to a 2LT matrix and use the Matrix Pencil and (auto-) Pairing[22] (MaPP) algorithm to retrieve the scattering points from the received signal, as will be explained in the sequel.

### 3.1 *2LT Toeplitz and Vandermonde Decomposition*

According to Eq. (3), (6), (7), (8), the received signal shown in matrix form in Eq. (4) can be decomposed as

$$S = YDZ. \tag{13}$$

where $Y, Z, D$ are given by

$$Y = [a(f_{x,1}), \ldots, a(f_{x,k})] \in \mathbb{C}^{N \times k},$$

$$Z = [a(f_{y,1}), \ldots, a(f_{y,k})] \in \mathbb{C}^{M \times k},$$

$$D = \text{diag}(d_1, \ldots, d_k) \in \mathbb{C}^{k \times k}. \tag{14}$$

Then, a $(2N-1) \times (2M-1)$ Toeplitz matrix is defined as

$$T = \begin{bmatrix} S_{-N,-M} & \cdots & S_{-N,0} & \cdots & S_{-N,M} \\ \vdots & & & & \vdots \\ S_{0,-M} & & \ddots & & S_{0,M} \\ \vdots & & & & \vdots \\ S_{N,-M} & \cdots & S_{N,0} & \cdots & S_{N,M} \end{bmatrix} \in \mathbb{C}^{(2N-1) \times (2M-1)} \tag{15}$$

based on which an $N \times N$ 2LT Toeplitz matrix is given by Ref. 18

$$\acute{T} = \begin{bmatrix} T_0 & T_{-1} & \cdots & T_{-(N-1)} \\ T_1 & T_0 & \cdots & T_{-(N-2)} \\ \vdots & \vdots & \vdots & \vdots \\ T_{N-1} & T_{N-2} & \cdots & T_0 \end{bmatrix} \in \mathbb{C}^{NM \times NM}, \tag{16}$$

where each submatrix $T_i$ is an $M \times M$ Toepiltz matrix defined as

$$T_i = \begin{bmatrix} s_{i,0} & s_{i,-1} & \cdots & s_{i,-(M-1)} \\ s_{i,1} & s_{i,0} & \cdots & s_{i,-(M-2)} \\ \vdots & \vdots & \vdots & \vdots \\ s_{i,M-1} & s_{i,M-2} & \cdots & s_{i,0} \end{bmatrix}. \tag{17}$$

Then, according to Theorem 2 in Ref. 19 and the Vandermonde decomposition, we can present $\acute{T}$ as

$$\acute{T} = V \Lambda V^*, \tag{18}$$

where $V = [c(f_1), \ldots, c(f_k)]$ is the Vandermonde matrix and $\Lambda = \text{diag}(\sigma_1, \ldots, \sigma_k)$ According to Theorem 1 in Ref. 19, if the rank of $\acute{T}$ is $k$, the $k$th-order Vandermonde decomposition of $\acute{T}$ is guaranteed by the MaPP algorithm. Then, the atomic $l_0$ norm of the received signal can be defined by the following non-convex rank minimization problem[19]

$$\min_{t, s_{n,m}, \acute{T}} \text{rank}(\acute{T}) \quad \text{subject to} \quad \begin{bmatrix} t & s_{n,m}^H \\ s_{n,m} & \acute{T} \end{bmatrix} \geq 0. \tag{19}$$

To solve Eq. (19), we use a convex relaxation to convert it to a Trace minimization problem provided that $\acute{T}$ admits a Vandermonde decomposition[19] as

$$\min_{t, s_{n,m}, \acute{T}} t + \text{Trace}(\acute{T}) \quad \text{subject to} \quad \begin{bmatrix} t & s_{n,m}^H \\ s_{n,m} & \acute{T} \end{bmatrix} \geq 0. \tag{20}$$

**3.2** *Reweighted Trace Minimization*

A smoothed approximation of Eq. (20) as a metric for $s_{n,m}$ is defined as[19]

$$\mathcal{L}^\mu(s_{n,m}) = \min_{t, \acute{T}} t + \ln|\acute{T} + \mu I| \quad \text{subject to} \quad \begin{bmatrix} t & s_{n,m}^H \\ s_{n,m} & \acute{T} \end{bmatrix} \geq 0. \tag{21}$$

where $\mu$ is the regularization factor. To solve Eq. (21), by defining $W$ as a weighting factor, we use the iterative algorithm RWTM, in which the $i$-th iteration of $\acute{T}$, denoted by $\acute{T}_i$, locally converges to an optimal point[19,]

$$\min_{t,\acute{T},s_{n,m}} t + \text{tr}\left|(T'_{i-1} + \mu_{i-1}I)^{-1}\acute{T}\right| \quad \text{subject to} \quad \begin{bmatrix} t & s_{n,m}^H \\ s_{n,m} & \acute{T} \end{bmatrix} \geq 0, \tag{22}$$

where the weighting factor is defined as

$$W = (T'_{i-1} + \mu_{i-1}I). \tag{23}$$

**3.3** *Proposed ISAR imaging Algorithm based on 2D-RWTM*

We propose an ISAR imaging method based-on 2D-RWTM algorithm to retrieve the frequencies of sparse scattering points in both range and cross-range directions. According to Fig.1, the target consists of a number of sparse scattering points. We assume that some transmitted pulses from different viewing angles of the target are not received. According to Eq. (9), the received signal vector consists of the RCS's of the scattering points and the 2D normalized complex atoms. Then, according to Eq. (15) and using the received signal vector in Eq. (9), we can form a Toeplitz matrix. Next, using Eq. (16), we get a 2LT Toeplitz matrix. Using Eq. (22), we propose a 2D-RWTM problem to retrieve the 2D-frequencies of scattering points, from which the locations of scattering points are subsequently estimated. As the number of scattering point increases, the resolution of ISAR images will enhance. The various steps of this method are summarized in Algorithm 1.

---

**Algorithm1: ISAR Imaging based on 2D-RWTM.**

**Input**: Received signal $S_R(f_m, \theta_n)$
**Output**: Retrieved scattering points of a target

1. Rewrite the received signal in matrix form using Eq. (3)
2. Use Eq. (7) and Eq. (8) to get 2D-atoms $a(f_{x,i})$, $a(f_{y,i})$
3. Find $s_{n,m} = \sum_{k=1}^{K} \sigma_k a(f_{x,i}) \otimes a(f_{y,i})$ in Eq. (6)
4. Form the Toeplitz matrix $T$ from $s_{n,m}$ in Eq. (15)
5. Form the 2LT Toeplitz matrix $\acute{T}$ from $T$ in Eq. (16)
6. Find the Vandermonde decomposition $\acute{T}$ in Eq. (18)
7. Defining the weighting factor Eq. (23)
8. Apply the iterative RWTM algorithm in Eq. (22)
9. Apply MaPP algorithm to retrieve scattering points

---

## 4 Experimental Results

We evaluate the proposed 2D-RWTM algorithm for an Airbus plane with 18 scattering points at a frequency of 9 GHz. Next, the MaPP algorithm is used to retrieve the locations of these points in both range and cross-range directions. To do so, 90 noise-free samples are randomly selected from a total of 196 complete data samples and the results of the 2D-RWTM algorithm are compared with those of the ANM algorithm. Also, to investigate the resolution of images and sparsity effect, the values of some factors such as the proximity of scattering points to each other and different weighting matrices are examined.

**4.1** *Weighting Matrix*

In this simulation, we investigate the effect of weighting matrices on retrieving the 2D-location of a target using the 2D-RWTM algorithm. To do so, we consider three different structures for the weighting matrices including; *a*) identity matrix, *b*) a matrix whose entries are one at a 128*128 central submatrix and zero in the rest as shown in Fig.2, and c) the entries are chosen randomly.

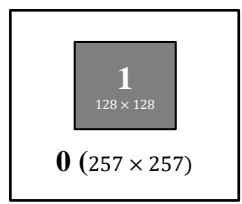

**Fig. 2**. Weighting matrix for Fig3. b.

The results are presented in Figs. 3a, b & c, where the red triangles show the actual location of the frequencies (scattering points) and the blue stars show the recovered frequencies. As seen, 18 scattering points with higher RCS than the others (shown by *) are retrieved, while the red triangles show the true scattering points. As seen, the 2D-RWTM algorithm with all the three weighting matrices generate acceptable results in retrieving the locations of scattering points.

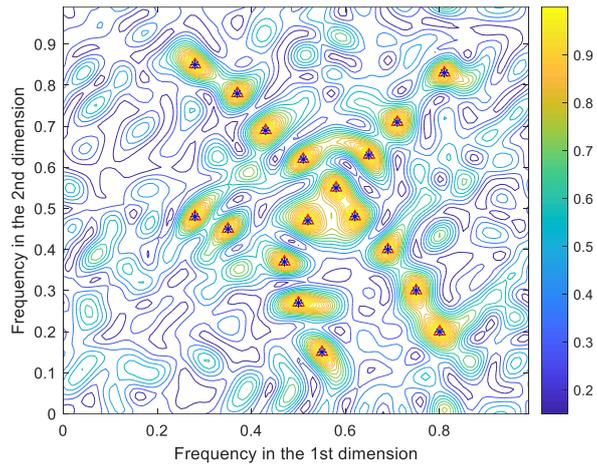

(a)

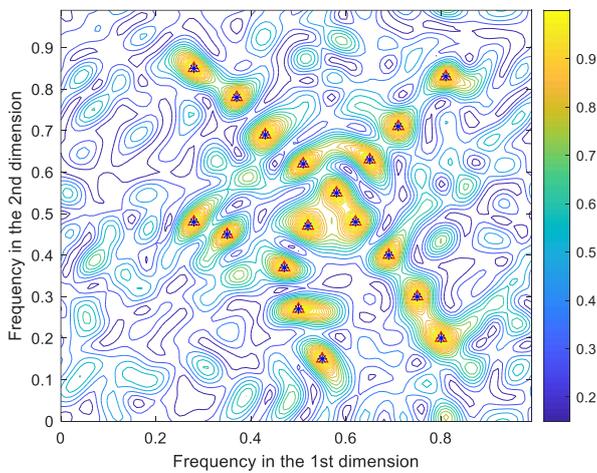

(b)

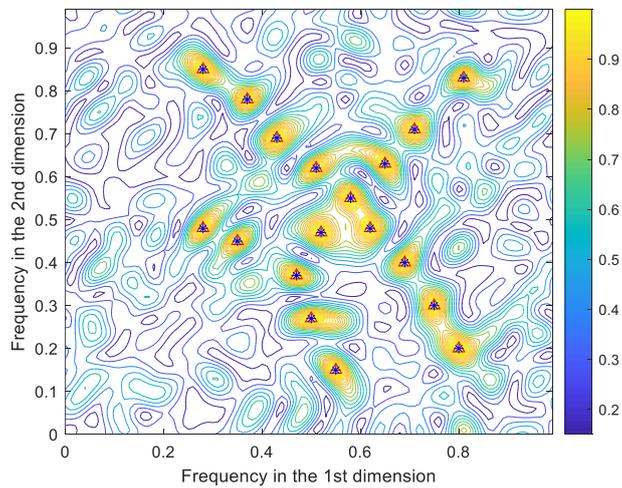

(c)

**Fig. 3.** Retrieval of 2D locations of scattering points of an Airbus plane using; a) identity weighting matrix, b) weighting matrix of ones in the center, and c) random weighting matrix.

**4.2** *Proximity of Scattering Points*

We examine the effect of proximity of the scattering points of a target. According to Ref. 19, the distance between two scattering points should not be less than $\Delta_f = \frac{0.45}{n_1}$, where $n_1$ is the number of samples in each direction. In other words, each scattering point in any direction should have such a distance from the other scattering points. Here, we consider $n_1 = 14$ and $\Delta_f = 0.032$. In ISAR imaging, this assumption is realistic and in practice, the points with high RCS can be as far apart as $\Delta_f$. For example, by considering 18 distinct scattering points, we easily retrieve the following 2D- frequencies:

$f_{1D} = [0.81, 0.71, 0.65, 0.43, 0.37, 0.28, 0.47, 0.75, 0.69, 0.8, 0.58, 0.28, 0.5, 0.55, 0.35, 0.51, 0.62, 0.52]$

$f_{2D} = [0.83, 0.71, 0.63, 0.69, 0.78, 0.85, 0.37, 0.3, 0.4, 0.2, 0.55, 0.48, 0.27, 0.15, 0.45, 0.62, 0.48, 0.47]$,

where $f_{1D}$ and $f_{2D}$ are the frequencies in the 1st and 2st dimension. In the first simulation, the distances of the above 18 scattering points satisfy the allowable value of $\Delta_f$. Then, using the 2D-RWTM and MaPP algorithms, we have successfully retrieved 18 two-dimensional frequencies, as seen in Fig. 4. Next, we consider 19 two-dimensional frequencies at

$f_{1D} = [0.80, 0.81, 0.71, 0.65, 0.43, 0.37, 0.28, 0.47, 0.75, 0.69, 0.8, 0.58, 0.28, 0.5, 0.55, 0.35, 0.51, 0.62, 0.52]$

$f_{2D} = [0.82, 0.83, 0.71, 0.63, 0.69, 0.78, 0.85, 0.37, 0.3, 0.4, 0.2, 0.55, 0.48, 0.27, 0.15, 0.45, 0.62, 0.48, 0.47]$

In either direction, the first two frequencies are very close to each other, so that the $\Delta_f$ condition is not met. In this case, the distance of scattering points from each other is 0.01, which is less than the allowable proximity $\Delta_f = 0.032$. The results in Fig.5 show that some of the frequencies are not recovered accurately.

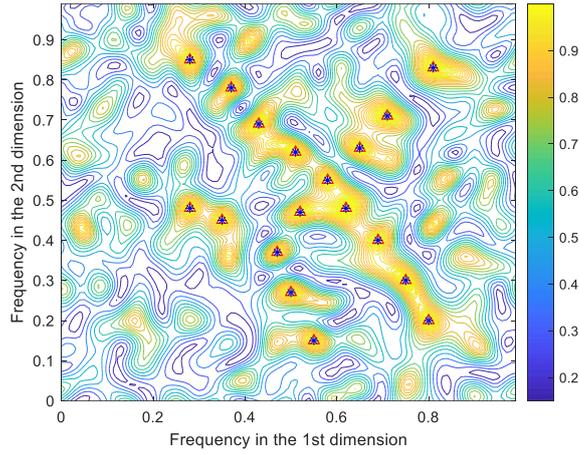

**Fig. 4.** Retrieval of 2D locations of scattering points, which satisfy the allowable distance of $\Delta_f$.

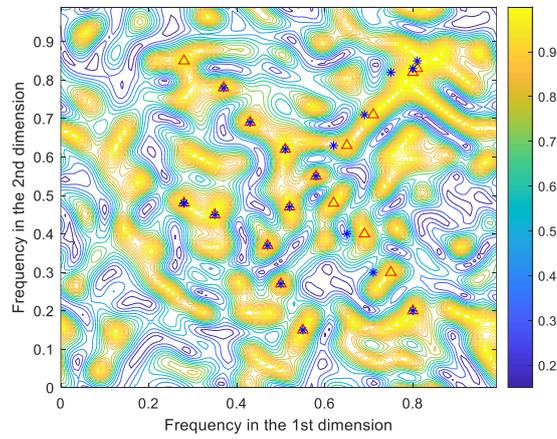

**Fig.5.** Retrieval of two-dimensional locations of scattering points, which do not satisfy the allowable distance of $\Delta_f$.

In the last experiment, the MSEs of 2D-RWTM and ANM are compared for a different number of samples. As seen in Fig.6, the MSE for the 2D-RWTM is lower than that of ANM, showing the outperformance of the proposed method for ISAR imaging.

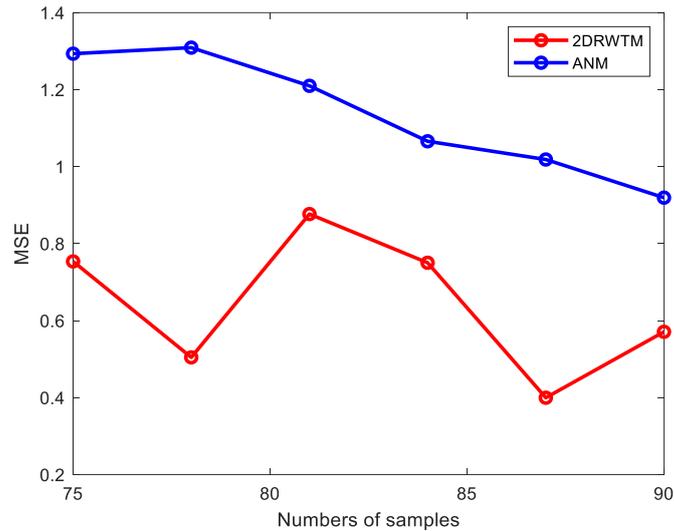

**Fig. 6.** Comparison of 2D-RWTM and ANM methods in ISAR imaging in terms of MSE based on the number of available samples

## 5 Conclusion

We proposed an ISAR imaging method for moving targets in order to increase the resolution of the retrieved images. In this scenario, we focused on developing an algorithm based on retrieving the sparse scattering points of targets, which generate stronger RCS's. Accordingly, we developed the 2D-RWTM algorithm by incorporating a practical assumption that the latter signals would be received incompletely. Simulation results showed that the 2D-RWTM can successfully retrieve the frequencies of scattering points of the targets in both ranges and cross-range. Also, this algorithm was effectively evaluated for different weighting matrices and generated a higher resolution in ISAR imaging with lower MSE's in comparison to the ANM.

.

## References

[1] V. C. Chen and M. Martorella, *Inverse Synthetic Aperture Radar Imaging: Principles, Algorithms and Applications.* Edison, NJ, USA: SciTech Publishing, (2014).

[2] C. Ozdemir, *Inverse Synthetic Aperture Radar Imaging with MATLAB Algorithms.* Hoboken, NJ, USA: Wiley, (2012).

**Mohammad H. Kahaei** received his PhD in signal processing from Queensland University of Technology, Brisbane, Australia, in 1998. Since 1999, he has been with the School of Electrical Engineering, Iran University of Science and Technology, where he is currently an associate professor and the head of the Signal and System Modeling Laboratory. His research interests include array signal processing with primary emphasis on compressed sensing, blind source separation, super resolution, tracking, and matrix completion.